# you can latex this if you have the aastex latex macro, Three compressed 
# uuencoded tar figures are added at the end of this file. To unpack them
# just strip off the main text and call resulting file, e.g., figs.uu
# then say csh figs.uu. if you are not on a unix machine, you should explicitly
# execute the commands: uudecode figs.uu; uncompress figs.tar.Z; 
# tar -xvf figs.tar. Or ftp the compressed postscript
# version using anonymous ftp on arcturus.mit.edu in /Preprints/cmb_approx.ps.Z

\documentstyle[12pt,aaspp]{article}
\begin{document}
\title{A TWO-FLUID APPROXIMATION FOR CALCULATING 
THE COSMIC MICROWAVE BACKGROUND ANISOTROPIES} 
\author{Uro\v s Seljak\altaffilmark 1}
\affil{Department of Physics, MIT, Cambridge, MA 02139 USA}
\begin{quote}
\altaffilmark{}
\altaffiltext{1}{Also Department of Physics, University of Ljubljana,
Jadranska 19, 61000 Ljubljana, Slovenia}
\end{quote}

\def\bi#1{\hbox{\boldmath{$#1$}}}
\begin{abstract}
We present a simple, yet accurate approximation for calculating the 
cosmic microwave background anisotropy power spectrum in adiabatic models. 
It consists of solving for the evolution of a two-fluid model 
until the epoch of recombination and then integrating over the sources to 
obtain the CMB anisotropy power spectrum. The approximation is
useful both for a physical understanding of CMB anisotropies, 
as well as for a quantitative analysis of cosmological models. Comparison 
with exact calculations shows that the accuracy is typically better than 
20 percent over a large range of angles and cosmological models, including 
those with curvature and cosmological constant. Using this 
approximation we investigate the dependence of the CMB anisotropies on the 
cosmological parameters. We identify six dimensionless parameters that uniquely 
determine the anisotropy power spectrum within our approximation. CMB 
experiments on different angular scales could in principle provide information 
on all these parameters. In particular, mapping of the Doppler peaks would 
allow an independent determination of baryon mass density, matter mass 
density and Hubble constant. 

\end{abstract}
\vskip 1cm
\section{Introduction}
Observations of fluctuations in cosmic microwave background (CMB)
can provide important constraints on cosmological 
models. Large angular scale ($>10^0$) observations  
probe the initial conditions, 
in particular the amplitude and the slope of primordial power spectrum 
(\cite{Smoot92}; \cite{Gorski94}; \cite{Wright94}). These scales could 
also provide information 
on the geometry and the matter content of the universe (\cite{Kofman85}; 
\cite{Kamio94};
\cite{Sugiyama94}). However, theoretical interpretation
of measurements on these scales
is complicated by cosmic variance and this intrinsically limits 
the accuracy with which these parameters can be estimated using 
large angular scale measurements alone.
While small scale measurements suffer less from cosmic 
variance, their interpretation is complicated by the microphysics 
during recombination and/or reionization. Theoretical models 
often give wildly different predictions for the anisotropy power 
spectra when the parameter values are only slightly changed, while 
some combinations of parameters seem to provide nearly 
identical spectra (e.g. \cite{Bond94}). 
The purpose of this Letter
is to clarify which combinations of physical parameters affect the 
CMB fluctuations and what are the physical processes that lead to 
these fluctuations. Our two-fluid model for adiabatic fluctuations, 
presented in \S  2, 
generalizes previous theoretical approximations that investigated CMB 
fluctuations in the limiting cases of large and small angles (\cite{SW66}; 
\cite{Jor93}). 
The model is accurate enough 
that it can be used for a quantitative analysis of various models, 
yet it is also simple enough that it can clearly separate between different 
physical processes that affect the CMB fluctuations. 
In \S  3 we use
this model to identify the physical parameters that can be 
determined using CMB measurements over a large angular range, extending
previous studies that were limited to a 
smaller range of angles and/or parameters (\cite{Bond94}; \cite{Kamio94b}; 
\cite{GS94}; \cite{Sugiyama94}; \cite{Muciaccia93}). 
In  \S 4 we present the conclusions.

\section{Method}

In this section we present a method for computing the CMB anisotropies
that is a generalization of analytic methods first introduced by Sachs
\& Wolfe (1966) 
and is based on the line-of-sight integration along the photon past 
light cone. 
We will assume that the photons and baryons are 
tightly coupled prior to recombination, which will allow a simple two-fluid
description of perturbations (\cite{Peebles70}). Our 
analysis will be restricted to the linear perturbation theory of 
adiabatic perturbations and 
we will neglect any possible tensor 
contributions. 
Different theoretical models will be compared 
using angular power spectrum of CMB anisotropies, $C_l=\langle 
\mid a_{lm} \mid^2\rangle$, where $a_{lm}$ is the multipole 
of temperature anisotropy $\Delta(\vec{n})=\sum_{l,m}a_{lm}Y_{lm}
(\vec{n})$ and $Y_{lm}
(\vec{n})$ is the spherical harmonic. Expected CMB anisotropy for a given 
experiment can be 
calculated from $\langle \Delta^2\rangle= \sum_{l\ge 2}(2l+1)W_lC_l/4\pi$,
where $W_l$ is the experiment window function.

The temperature fluctuation $\Delta(\vec{n})$
in the direction $\vec{n}$ can be expressed as a line-of-sight integral,
\begin{equation}
\Delta(\vec{n})=\int_0^{\tau_0}
[\dot{\mu}(\phi+{\delta_{\gamma}\over 4}
+\vec{n}\cdot \vec{v_b})+
2\dot{\phi} ]e^{-\mu} d\tau.
\label{bolt-sol}
\end{equation}
Here $\tau$ is the conformal time with the value $\tau_0$ today, 
$\phi$ is the gravitational potential,
$\delta_{\gamma}$ is the photon density perturbation and $\vec{v_b}$ is the 
electron velocity. 
We introduced the Thomson opacity along the past light cone
$\mu(\tau)=\int_{\tau}^{\tau_0}\dot{\mu}(\tau')d\tau'$ with
$\dot{\mu}=ax_en_e\sigma_T$, where $a$ is the expansion factor, 
$x_e$ the ionization fraction, $n_e$ the
electron number density and $\sigma_T$ the Thomson cross section.
The above expression is written in 
the gauge invariant formalism (\cite{Kodama84}).  
We neglected the anisotropic stress contribution and the
unobservable monopole contribution arising from the local gravitational
potential.
In the limit of infinitely thin LSS
the function $\dot{\mu}e^{-\mu}$ approaches to a Dirac delta-function  
$\delta(\tau-\tau_{rec})$, where $\tau_{rec}$ denotes the conformal
time at recombination. Equation \ref{bolt-sol} then reduces to 
(Kodama \& Sasaki 1984)
\begin{equation}
\Delta(\vec{n})=\phi(\tau_{rec})+{\delta_{\gamma}(\tau_{rec})\over 4}+
\vec{n}\cdot \vec{v_b}(\tau_{rec})
+2\int_{\tau_{rec}}^{\tau_0}\dot{\phi}(\tau) d\tau.
\label{bolt-sol-rec}
\end{equation}

The velocity term in equation
\ref{bolt-sol-rec} can be 
rewritten using the linear theory approximation $\vec{v_b}=\vec{\nabla} \psi_b$
into $\vec{n}\cdot\vec{v_b}=\partial\psi_b / \partial r$, where $r$ is 
the radial coordinate. We may  
decompose the sources in equation \ref{bolt-sol-rec} into an
orthonormal spherical basis set with Fourier amplitudes $\phi(k)$, 
$\delta_\gamma(k)$ and $v_b(k)$, where $v_b(k)=i\psi_b(k)/ k$. 
After the angular and ensemble averaging 
we obtain the following expression for the multipole moments
(\cite{Kodama84}),
\begin{eqnarray}
C_l&=&4\pi \int_0^{\infty}k^2P(k)T(k)D_l^2dk \nonumber \\
D_l&=&(\phi+ {\delta_\gamma \over 4})j_l(k
\tau_0-k\tau_{rec})+vj_l'(k\tau_0-k\tau_{rec})+2\int_{\tau_{rec}}
^{\tau_0}d\tau
j_l(k\tau_0-k\tau)\dot{F}(\tau),
\label{cl}
\end{eqnarray}
where $j_l$ is the spherical Bessel functions and $j_l'$ its
derivative. All the perturbed quantities
are evaluated in $k$-space at $\tau_{rec}$.
$P(k)$ denotes the primordial power spectrum of potential $\phi$, 
usually expressed as a power law $P(k) \propto k^{n-4}$. For later purpose 
we introduced the function $T(k)$, which incorporates the damping effects. 
In the case of a non-flat universe the functions $j_l$ need to be
substituted by their appropriate generalizations. This is 
unimportant for $l >\Omega_0^{-1}\vert 1-\Omega_0\vert^{1/2}$ 
and the results
presented here will be valid for non-flat universes,
provided that the relation between angles and
physical sizes is expressed using angular size distances.
Last term in equation \ref{cl} gives the so-called
integrated Sachs-Wolfe (ISW) contribution. The time dependence of the 
potential is denoted with $\dot{F}(\tau)$, where $F(\tau)=\phi(\tau)/
\phi(\tau_0)$. 
This term vanishes in a flat, matter dominated $\Omega_m=1$
universe, but is present in the case of a vacuum energy dominated universe
(\cite{Kofman85}),
curvature dominated universe (\cite{Kamio94}) or when the universe is in
transition epoch from being radiation to being matter 
dominated (\cite{Kodama86}). 

To calculate the anisotropy power spectrum 
we need to evaluate the source contributions
in equation \ref{cl} at the epoch of recombination. 
The photon evolution equations in $k$-space are given by
(\cite{Ma94}; \cite{Wilson81}; \cite{Bond84})
\begin{equation}
\label{photon2}
     \dot{\delta}_\gamma = -{4\over 3}kv_\gamma
        +4\dot{\phi}, \ \ \ \ 
     \dot{v}_\gamma = 
{k \delta_\gamma \over 4}
        + \dot{\mu} (v_b-v_\gamma) 
         + k \phi.
\label{photon}
\end{equation}
We also need the evolution equations for baryon and CDM perturbations, 
\begin{eqnarray}
        \dot{\delta}_b &= -kv_b + 3\dot{\phi}, \ \ \ \  
        \dot{v}_b &= -{\dot{a}\over a}v_b
         + {4 \bar\rho_\gamma\over 3\bar\rho_b}
         \dot{\mu} (v_\gamma-v_b) + k\phi\, \nonumber \\
        \dot{\delta_c} &= -kv_c + 3\dot{\phi},\ \ \ \  
        \dot{v}_c &= - {\dot{a}\over a}\,v_c+k\phi ,
\label{cdmb}
\end{eqnarray}
where $\bar\rho_b$ and $\bar\rho_\gamma$ are  
the baryon and photon
mean densities, respectively. 

The energy and momentum constraint
equations give the equations for $\phi$ and $\dot{\phi}$,
\begin{equation}
\phi=-{4\pi G a^2\over k^2}
(\rho+{3 \dot{a} f\over ak}),\ \ \ \  \dot{\phi}=-{\dot{a}\over a}\phi
+{4\pi Ga^2f \over k},
\label{constr}
\end{equation}
where $
\rho=(\bar\rho_\gamma+\bar\rho_\nu) \delta_\gamma + \bar\rho_b \delta_b
+\bar\rho_c\delta_c$ and $f={4\over 3} (\bar\rho_\gamma+\bar\rho_\nu) v_\gamma
+\bar\rho_b v_b+\bar\rho_c v_c$. Here $\bar\rho_\nu$ and $\bar\rho_c$ are 
the neutrino and CDM mean densities, respectively.
We replaced neutrino density and 
velocity perturbations with the corresponding photon perturbations. 
This 
becomes invalid on small scales due to the free-streaming of neutrinos, 
but does not affect significantly 
the final results. We also neglected the anisotropic shear and 
the possible curvature terms.

The above equations are 
supplemented by the Friedmann equation, which at early times (when a 
possible cosmological constant or curvature term can be neglected) is 
given by
\begin{equation}
({\dot{a}\over a})^2={8\pi G a^2 \over 3}(\bar\rho_\gamma+\bar\rho_\nu+
\bar\rho_b+\bar\rho_c) 
.
\label{friedmann}
\end{equation}
The solution to this equation is
\begin{equation}
y\equiv {a \over a_{eq}}=(\alpha x)^2+2\alpha x,\ \ \ \  x=\left
 ({\Omega_{m} \over
a_{rec}}\right )^{1/2}{H_0 \tau \over 2}\equiv {\tau \over \tau_{r}},
\label{fried-sol}
\end{equation}
where 
$a_{eq}=
(\bar\rho_\gamma+\bar\rho_\nu)/(\bar\rho_b+\bar\rho_c)\approx 4.2\times
10^{-5}\Omega_{m}^{-1}h^{-2}$ (assuming three flavors of massless neutrinos),
$a_{rec}^{-1}\approx 1100$ for the standard recombination,
$\alpha^2\equiv a_{rec}/a_{eq}$, 
$\Omega_{m}=\Omega_{b}+\Omega_{c}$ 
is the value of matter density today in units 
of critical density and  
$h$ is the value of Hubble constant today in units of 100km/s/Mpc.

We will now assume the tight coupling limit $\mu\gg 1$, 
which is a good approximation  
on scales larger than the Silk damping scale (\cite{Silk68},
\cite{Peebles70}). 
In this case the photons and
baryons are coupled into a single fluid with $\delta_b={3 \over 4}
\delta_\gamma$ and $v_b=v_\gamma$. The above equations rewritten
in terms of dimensionless time $x$ and dimensionelss wavevector $\kappa=
k\tau_{r}$ become
\begin{eqnarray}
\dot{\delta}_c=&-\kappa v_c+3\dot{\phi} \,, \ \ \ \ \
&\dot{v}_c=-\eta v_c+\kappa \phi \nonumber \\
\dot{\delta}_\gamma=&-{4 \over 3}\kappa v_\gamma +4 \dot{\phi} \,, \ \ \ \ \
&\dot{v}_\gamma=
({4 \over 3}+y_b)^{-1} 
\left[-\eta y_bv_\gamma +
{\kappa \delta_\gamma \over 3}
+ \kappa \phi ({4 \over 3}+y_b) \right] 
\nonumber \\
\phi=&-{3\over 2} (\eta/\kappa)^2(\delta+3\eta v /\kappa)  \,,\
\ \ 
&\dot{\phi}=-\eta \phi+ {3 \eta^2 v \over 2 \kappa} \nonumber \\
\delta =&{\delta_\gamma[1+{3\over 4}(y-y_c)]+y_c\delta_c 
\over 1+y}\,, \ \ \ \
&v ={v_\gamma({4 \over 3} +y-y_c)+y_cv_c \over 1+y},
\label{eqs}
\end{eqnarray}
where the derivatives are taken with respect to $x$, $y_b\equiv 
{\bar\rho_b\over \bar\rho_\gamma}=[1+{3\times 7\over 8}({4
\over 11})^{4/3}]{\Omega_{b} \over
\Omega_{m}}y= 1.68 {\Omega_{b} \over
\Omega_{m}} y$,
$y_c={\Omega_{c} \over \Omega_{m}}y=(1-{\Omega_b\over \Omega_m}
/1.68)y$ 
and $\eta=2\alpha(\alpha x+1)/ (\alpha^2 x^2+2\alpha x)$. 
Equations \ref{eqs} 
are a coupled system of 4 first order differential equations.\footnote[2]{
In actual numerical implementation of these equations we find that for 
a stable numerical integration it is better to compute $\phi$ using its 
time evolution in equation \ref{eqs}, rather than computing it from 
the sources.} The appropriate initial conditions at $x<<1$ (when the 
universe is radiation dominated) and 
$\kappa \eta <<1$ (when the mode is larger than the Hubble sphere radius) are 
\begin{eqnarray}
\phi=1\,,\ \ \delta_\gamma=-2\phi(1+{3y \over 16})\,,\ \  \delta_c={3 \over 4} 
\delta_\gamma \nonumber \\
v_\gamma=v_c=-{\kappa \over \eta}
\left[{\delta_\gamma \over 4}+ {2 \kappa^2(1+y)\phi \over 
9\eta^2({4\over 3}+y)}\right].
\label{init}
\end{eqnarray}

The above equations need to be evolved until 
$x_{rec}=[(\alpha^2+1)^{1/2}-1]/\alpha$.  
The temperature anisotropy expressed with the dimensionless
variables is given by
\begin{equation}
C_l=4\pi A \int_0^{\infty}\kappa^nT(\kappa)d\ln \kappa
\left[\left(\phi+{\delta_\gamma\over 4}+2\Delta \phi)\right)
j_l(\kappa x_0
)+v_\gamma j_l'(\kappa x_0)\right]^2,
\label{clfinal}
\end{equation}
where $x_0$ is the angular distance to the LSS in units
of $\tau_r$ and 
we assumed $P(k)=Ak^{-3}\kappa^{n-1}$. The term $\Delta \phi=
[2-8/y(x_{rec})+16x_{rec}/y^3(x_{rec})]/10y(x_{rec})$ 
arises from the ISW effect 
due to the potential varying with time during the transition period 
from the radiation dominated to the matter dominated universe
(\cite{Kodama86}). 
For simplicity we dropped the ISW contribution
from possible
curvature or cosmological constant, which is 
only important at the lowest values of $l$ ($l<10$). 

The damping transfer function $T(\kappa)$ is approximately unity
for low values of
$l$ ($l <200)$, but gradually decreases afterwards. Its main contributions come
from the Silk damping and from the finite width of LSS.
The first effect can be calculated analytically 
by expanding equations \ref{photon} and \ref{cdmb} to  
second order in $\dot{\mu}$ and neglecting the effects of gravity and 
expansion. In the matter-dominated era one obtains
$T(\kappa)\propto exp(-2\kappa^2x_s^2)$ 
(\cite{Fug90}),
where $x_s$ is the Silk damping scale in units of 
$\tau_r$,
$x_s=0.6\Omega_m^{1/4}\Omega_b^{-1/2}a_{rec}^{3/4}h^{-1/2}$.
Second effect can be analytically estimated
by performing the line-of-sight integral in
equation \ref{bolt-sol} in the limit where the sources are slowly changing 
over the timescale on which the visibility function $\dot{\mu}e^{-\mu}$ is
non-negligible (\cite{Jor93}). Visibility function can be
approximated as a gaussian $(2\pi\sigma^2x_{rec}^2)^{-1/2}exp[-(x-x_{rec})
^2/2(\sigma x_{rec})^2]$, 
where for standard recombination $\sigma \approx 0.03$.\footnote[3]{Both 
$\sigma$ and $a_{rec}$ are weakly dependent on cosmological parameters. 
Moreover, the ISW visibility function differs from the
Thomson scattering visibility function, which leads to a different damping 
of ISW term. Both
effects will be neglected here.} In the limit
$\kappa_0 x_0>>l$ (where $\kappa_0$ is the wavevector which gives the dominant
contribution to $C_l$), 
we obtain the damping factor $exp(-\kappa^2\sigma^2x_{rec}^2)$. Therefore, 
the damping effects can be written as 
\begin{equation}
T(\kappa) \approx e^{-\kappa^2 (2x_s^2+\sigma^2x_{rec}^2)}.
\label{transfer}
\end{equation}
This works reasonably well for the standard ionization history. 
Note however that 
if the limit $\kappa_0 x_0>>l$ is not satisfied, then the
damping due to the
finite thickness of LSS is
not exponential, but is proportional to $\kappa^{-1}$. This will be the
case, for example, in reionized models.

\begin{figure}[p]
\includegraphics{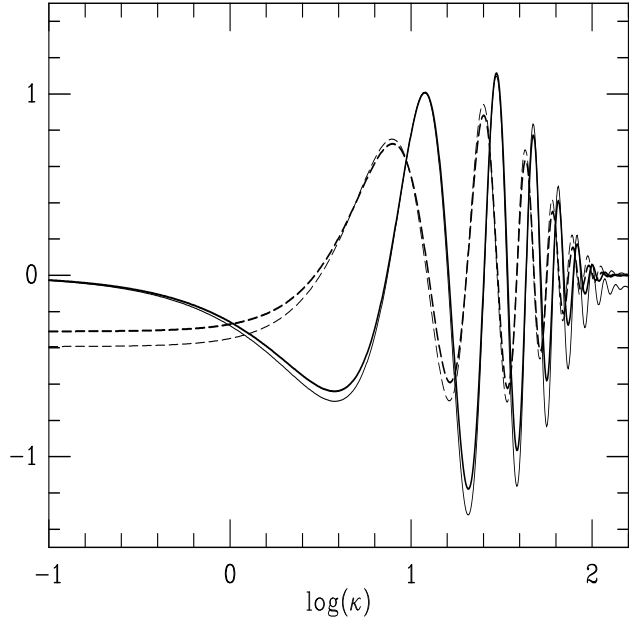}
\vspace*{7.5cm}
\caption{Comparison between our approximation (thick lines) 
and exact solution (thin lines) for 
$v_\gamma$ (solid lines) and $\phi+\delta_\gamma/4$ (dashed lines) 
as a function of $\kappa$.
Silk damping has been included according to the expression in the
text. Parameter values are $\Omega_b=0.05$, $h=0.5$ and $\Omega_m=1$.}
\label{fig1}
\end{figure}

Equations \ref{eqs}-\ref{transfer}  
are all is needed to 
evaluate the temperature fluctuations. Although 
equations \ref{eqs}
cannot be solved analytically in general, they have analytic
solutions in the limits of small and large $\kappa$.  
In the first limit where the modes are larger than the Hubble sphere radius
the amplitude of perturbations at a given time is a constant (figure
\ref{fig1}). This  
gives the standard Sachs-Wolfe expression for CMB fluctuations, as can 
be verified by evolving the initial conditions in equations \ref{init} 
into the matter-dominated era  
and neglecting the velocity term in
equation \ref{clfinal}. In the second limit (large $\kappa$) the
equations can be solved using the WKB approximation 
and the solution is given by the acoustic oscillations
of photon-baryon plasma
(\cite{Jor93}, \cite{Pad93}). 
In the intermediate regime, which 
is of main interest for us, the
equations need to be solved numerically,  
but the physics can be well 
understood by the two limits above. 
As shown in figure \ref{fig1}, 
equations \ref{eqs}
give an excellent approximation to the exact results over a 
large range of wavevector $\kappa$. 

Although the above system of equations is particularly useful for 
the standard recombination scenarios, one can also use it to calculate 
anisotropy power spectrum in reionized models. As one can see from 
equation \ref{bolt-sol}, the 
primary fluctuations will be suppressed by a factor $exp[-\mu(x_{rec})]$,
where $\mu(x_{rec})$ is Thomson opacity at recombination. In addition, 
there will be secondary fluctuations generated at the new last-scattering
surface, which can be calculated using the same method as above, except
that one needs to replace $v_\gamma$ with $v_c$ in the regime where the 
Compton drag is negligible. 
Since the modes larger than the Hubble sphere radius 
at the new LSS do not evolve in the matter-dominated regime
this simply regenerates the Sachs-Wolfe expression for low values of l. 
On smaller scales
the thick new LSS damps the secondary fluctuations and in many  
scenarios these become negligible (although on arcminute scales  
the second-order terms may become important, \cite{Vishniac87}).
Provided that one is interested in degree angular scales, then the 
effect of reionization is to suppress the fluctuation power spectrum 
relative to the large scales by a factor of $exp[-2\mu(x_{rec})]$.

\section{Results}
\begin{figure}
\includegraphics{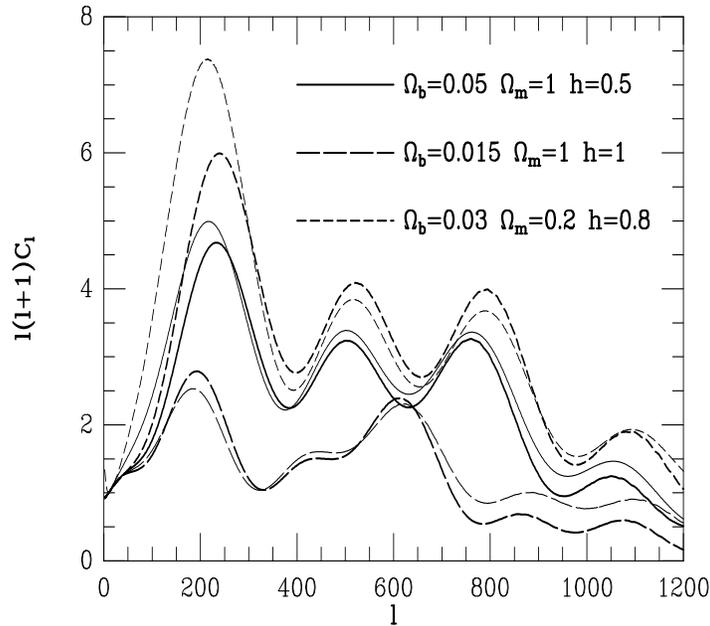}
\vspace*{7.5cm}
\caption{Comparison of anisotropy power spectra 
between our approximation (thick lines) 
and exact solution (thin lines) 
as a function of multipole moment $l$
for several different cosmological 
models. The spectra in all figures are normalized relative to $C_{10}$.}
\label{fig2}
\end{figure}

The comparison between our approximation and the exact solutions of perturbed 
Boltzmann equation (Seljak \& Bertschinger 1994; Seljak 1994) 
is presented in figure 
\ref{fig2} for several cosmological models. All of the multipole moments are 
normalized relative to $C_{10}$, which is approximately
fixed by the COBE experiment and where the curvature effects and 
ISW effects due to $\Omega_m \neq 1$ can be neglected. 
One can see that the agreement
is excellent over a large range of $l$. The deviations at large 
$l$ arise because of improper treatment of damping effects,
while the deviations at small $l$ can be attributed to the 
neglection of neutrino anisotropic shear.
Another effect that introduces small deviations is the dependence of $a_{rec}$  
on cosmological parameters, which slightly offsets the position 
of the peaks. Nevertheless, our approximation
correctly predicts the positions and  
amplitudes of Doppler peaks (also called acoustic or Sakharov oscillations)
with a 20 \% accuracy over most of parameter range.

Given the high accuracy of our model we may now investigate how the 
anisotropy power spectrum depends on the cosmological model. Our goal
is to identify the parameters that can be determined using the CMB 
measurements and to give a physical understanding of how they affect
the anisotropies.
In equations \ref{eqs} the free parameters are $\alpha=
21.5 \Omega_{m}^{1/2}h$ and ${\Omega_{b}\over \Omega_{m}}$. We can
replace them with physically 
more relevant parameters $\Omega_mh^2=(\alpha/21.5)^2$ and $\Omega_bh^2=
(\alpha/21.5)^2{\Omega_b\over \Omega_m}$. In addition
to these two we have the parameters $\mu(x_{rec})$,
$n$, $x_0$ and $x_s$. These six parameters will uniquely determine all the 
CMB power spectra within our approximation.
Of the parameters above, 
$n$, $\mu(x_{rec})$ and $x_s$ all suppress the power
on small
scales relative to large scales (for $n<1$). The suppression 
is different in the three cases, being proportional to $\kappa^{n-1}$ 
(neglecting the possible tensor contribution),
$exp[-\mu(x_{rec})]$ and $exp(-2\kappa^2x_s^2)$, respectively. This, in 
principle, allows  
to separate the different suppression effects and to determine the three 
parameters separately (but see \cite{Bond94}).
Note that these parameters do not change the positions of the 
Doppler peaks, only their amplitude. Since the effects of these 
parameters are physically transparent, 
we will restrict in the following to the case of $n=1$ 
and $\mu(x_{rec})=0$. 
Moreover, Silk damping is important only for large values of $l$
and it can be neglected if one concentrates
on the first few Doppler peaks.
We are thus left with $x_0$, $\Omega_mh^2$ and
$\Omega_bh^2$, which  
uniquely determine positions  
of the Doppler peaks. 

\begin{figure}[p]
\includegraphics{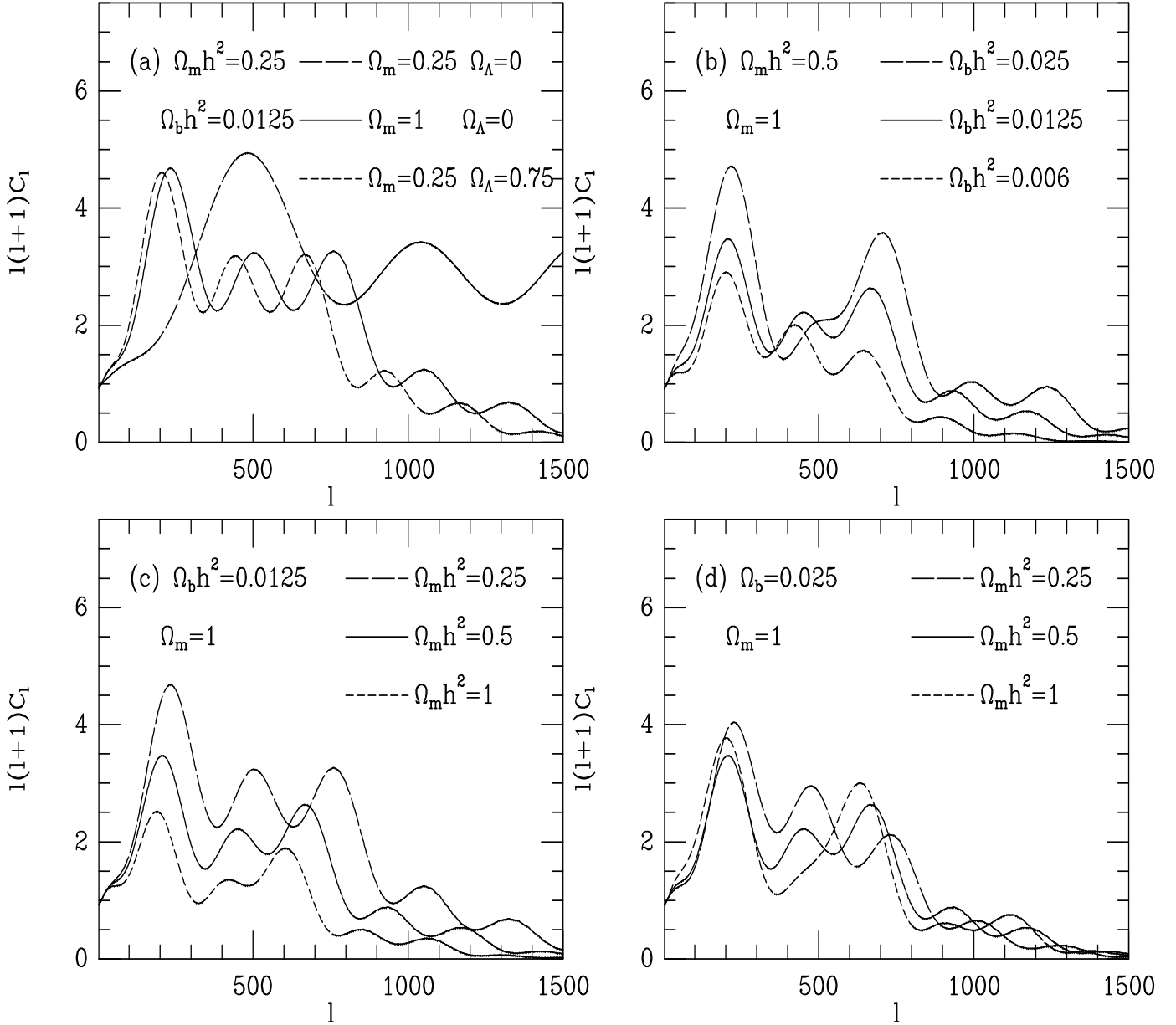}
\vspace*{17cm}
\caption{Anisotropy power spectra as a function of multipole moment $l$ for 
different cosmological models. In (a) curvature and cosmological constant
dominated models with 
$\Omega_m=0.25$ are compared to $\Omega_m=1$ model. 
In (b) $\Omega_mh^2$ is fixed at 0.5 and $\Omega_b
h^2$ is varying, whereas  in (c) $\Omega_b
h^2$ is fixed at the nucleosynthesis value and $\Omega_mh^2$ is varying. 
In (d) $\Omega_b/\Omega_m$ is fixed and $\Omega_mh^2$ is varying. 
In (b), (c) and (d) $\Omega_m=1$. In all 
cases varying the parameter changes the pattern of Doppler peaks.}

\label{fig3}
\end{figure}

Position of the first Doppler peak is determined by the 
angular size of the Hubble 
sphere radius at decoupling, which, expressed in terms of our variables is
given by $(1+\alpha^{-2})^{1/2}x_0$. 
Since $\alpha>>1$ for typical values of $\Omega_mh^2$, position of the
first Doppler peak mainly depends on $x_0$.
Assuming $x_{rec}<<x_0$ we have $x_0=(\Omega_{m}a_{rec})^{-1/2}$ for 
the model with negligible
cosmological constant and $x_0 = \Omega_{m}^{0.09}a_{rec}^{-1/2} $ 
for the model with negligible curvature. The
latter result shows that the first Doppler peak only weakly depends
on $\Omega_{\lambda}=1-\Omega_m$. 
This is because the angular size distance at large 
redshifts scales
with $\Omega_m$ in approximately the same way as does the Hubble 
sphere radius at decoupling (\cite{Vittorio85}; \cite{GS94}). 
This is, however, not true in general and the position
of the first Doppler peak depends on $\Omega_m$ and 
$\Omega_{\lambda}$ when both 
curvature and cosmological constant are important. 
The value of $l$ at 
which the maximum of the first Doppler peak lies is given approximately by 
$6x_0$. 
Figure \ref{fig3}a compares $C_l$'s of curvature and cosmological 
constant dominated models with those of
$\Omega_m=1$ model at fixed values of $\Omega_mh^2$ and $\Omega_bh^2$.
One can see that the position of the first Doppler peak 
can accurately determine $\Omega_m$ in curvature dominated universe
(\cite{Kamio94b}),
but cannot precisely 
determine $\Omega_m$ in cosmological constant dominated universe 
(\cite{Bond94}; \cite{GS94}).
However, even in this model the positions of secondary 
Doppler peaks are already significantly displaced relative to each other
when $\Omega_m$ changes from 0.25 to 1. This  
would thus allow independent determination of $\Omega_m$ even in
cosmological constant dominated universe, once $\Omega_mh^2$ and
$\Omega_bh^2$ are known (see below).
 
The dependence of the Doppler peak 
positions and amplitudes on $\Omega_bh^2$ and 
$\Omega_mh^2$ is more complicated, since both parameters appear 
in the evolution equations \ref{eqs} and change the properties of 
acoustic oscillations. Moreover, the two parameters enter into the 
equations differently and 
have different physical effects: $\Omega_bh^2$
is related to the properties of photon-baryon plasma and determines
its effective sound velocity at recombination, whereas $\Omega_mh^2$
is related to the time evolution of the expansion factor, since it
determines the epoch of matter-radiation equality. This means that one cannot
expect the anisotropy spectra to remain invariant under a certain combination of
the two parameters and both 
$\Omega_bh^2$ and $\Omega_mh^2$ are required for a complete description 
of the Doppler peaks. Figures \ref{fig3}b and \ref{fig3}c 
show how the Doppler peaks 
change when one of the two parameters is changing while the other is 
held fixed. If one concentrates on the first Doppler peak then 
it is not possible to determine the two parameters simultaneously, 
since both increasing $\Omega_bh^2$
and decreasing $\Omega_mh^2$ lead to an increase in the first Doppler
peak. The physical mechanisms that lead to this are 
different: while an increase in $\Omega_bh^2$ increases the amplitude of
the first wave in $v_\gamma$ and $\phi+\delta_\gamma/4$ (fig.
\ref{fig1}), a decrease in $\Omega_mh^2$ also leads to an increased ISW
contribution. 
Once the secondary peaks are observed as well, then different 
effects of the two parameters become significant and allow one to 
determine the two parameters simultaneously (figs.
\ref{fig3} b,c,d). Figure \ref{fig3}d shows how
changing $\Omega_mh^2$
at a fixed value of ${\Omega_b \over \Omega_m}$ 
affects the multipole moments. Again, since
the epoch of matter-radiation equality
is changing with $\Omega_mh^2$, one does not expect the 
multipole moments to remain unchanged and both  
our approximation and exact calculations confirm
this. Therefore, changing $h$ at a fixed $\Omega_b$ and
$\Omega_m$ changes the anisotropy power spectrum, 
contrary to some recent claims (\cite{Bond94}; \cite{GS94}).

\section{Discussion}

The approximation for calculating anisotropy power spectrum 
presented here is a generalization of the Sachs-Wolfe approximation,
which itself is only valid on scales larger than the Hubble sphere radius at 
recombination. By modelling the cosmological perturbations as a 
two-component fluid plasma we extended this approach
to all scales. 
The approximation is useful both for developing 
the physical understanding of processes that affect CMB fluctuations,
as well as for a quantitative prediction of multipole moments for various 
cosmological models. 
The main approximations used in our model are a two-fluid approximation,
neglection of anisotropic shear, a simplified treatment of 
Thomson scattering effects, neglection of curvature effects and neglection 
of vector and tensor contributions. None of these assumptions is essential
for the method and one can generalize the approach presented here
to obtain exact results (Seljak 1994). This will lead to
a computationally more demanding system of equations, but the main physical
effects that lead to the creation of Doppler peaks will still be determined
by the equations presented in this Letter.

By rewriting the equations in their dimensionless
form we identified all the dimensionless parameters that 
affect the anisotropy power spectra. 
Measurements of CMB fluctuations can only determine these parameters.
For example, neutrinos enter into our equations 
indirectly through the Friedmann equation \ref{friedmann} and through
the energy-momentum constraint equations \ref{constr}. 
The presence of a massive neutrino
only weakly changes these equations and the resultant multipole moments are 
almost indistinguishable from the ones with the massless neutrino. 
Therefore, the question of whether neutrino has a mass has little hope to
be answered using the CMB measurements.

The most interesting aspect of the CMB power spectra is the peculiar 
pattern of the Doppler peaks, which allows a simultaneous
determination of $\Omega_bh^2$ and $\Omega_mh^2$. 
This would provide an independent test of nuclesynthesis prediction of 
$\Omega_bh^2$ (e.g. \cite{Walker91}) 
and would also constrain the parameter space on $\Omega_m$ 
and $h$.
In addition, the position of the first 
Doppler peak determines $\Omega_m$ in curvature dominated model. In 
cosmological constant dominated model the position of the first Doppler
peak does not allow one to determine $\Omega_m$ accurately, but positions 
of secondary Doppler peaks could be used to constrain 
$\Omega_m$ (although for accurate determination 
exact calculations should be used in this case). 
Another way to break the degeneracy 
between $\Omega_m$, $\Omega_b$ and $h$
is to determine the Silk damping scale $x_s$, 
which depends only on these three parameters and cannot be expressed 
as a combination of $\Omega_bh^2$ and $\Omega_mh^2$.
This would require a separation of Silk damping from the 
damping due to the finite thickness of LSS in reionized models and from 
the $n<1$ suppression of small scales relative to large scales 
(including the possible tensor contribution). This is possible, 
because the three effects suppress the small scale power differently. 
Thus, a combination of CMB measurements over a large range of angles could 
be used to separately 
determine the baryon mass density, matter mass density and
the Hubble constant value.

I would like to thank Ed Bertschinger for useful discussions.
This work was supported by grants NSF AST90-01762 and NASA NAGW-2807.

\end{document}